\documentclass[twocolumn,floatfix,eqsecnum,aps,pra,nofootinbib]{revtex4-2}
\usepackage{graphicx}
\usepackage{amsmath}
\usepackage{amssymb}
\usepackage{bm}
\usepackage{hyperref}

\usepackage{color}

\begin{document}

\author{C. W. J. Beenakker}
\title{Casimir effect of symmetrically gapped chiral fermions}

\affiliation{Instituut-Lorentz, Universiteit Leiden, P.O. Box 9506, 2300 RA Leiden, The Netherlands}

\date{November 2026}

\begin{abstract}
The Casimir effect (the emergence of a force from confined quantum vacuum fluctuations) is a sensitive probe of the reflection of massless particles by gapped boundaries. Here we study the effect in the context of symmetric mass generation (SMG), the opening of an excitation gap by interactions without the breaking of a symmetry. The reflection of massless chiral fermions by an SMG-gapped region then involves composite many-body excitations, in order to preserve the ${\rm U}$(1) symmetry of charge conservation. We apply a recently developed sine-Gordon bosonization of the Wang \& Wen model of SMG (the one-dimensional charge 3-4-5-0 chiral fermion model) to compute the Casimir force between two gapped regions separated by a gapless interval of length $L$. For large $L$ the force decays as $1/L^2$, with a coefficient that depends on the conserved charges of the junction. For small $L$ it is governed by the Luttinger parameter $K\in(0,2/5)$ of the repulsive Hubbard interaction: When $K>1/5$ the force increases $\propto L^{2-10K}$ as $L\rightarrow 0$, while for $K<1/5$ it reaches a finite value at contact, which we obtain directly from the bulk energy density of the sine-Gordon model. At $K=1/5$ the model maps onto a superconducting Josephson junction, with the conserved charges playing the role of the conserved spin of the superconductor and the phase of the gapping interaction in the role of the superconducting phase. This mapping gives the full dependence of the Casimir force on $L$, including a logarithmic increase in the short-junction regime.
\end{abstract}
\maketitle

\section{Introduction}
\label{sec:intro}

The Casimir effect is the confinement dependence of the vacuum energy of a quantum field, measurable via a force that attracts or repels the boundaries \cite{Mos97}. While originally discovered for bosonic (electromagnetic) fields between conductors \cite{Cas48}, fermionic analogues have been studied in particle physics (confined quarks \cite{Mil80,Baa83}) and in condensed matter physics. In the latter context the filled Fermi sea plays the role of vacuum and the Casimir force is mediated by gapless quasiparticle excitations (Dirac fermions in graphene or topological insulators \cite{Zha08,Shy09,Rod14,Ish21,Lu21}, Bogoliubov quasiparticles in a superconductor \cite{Kri04,Bul05,Bee23,Lev25}). The Casimir effect of collective excitations (solitons) in a Luttinger liquid \cite{Faz95,Mas96,Aff00,Cau02,Rec05,Fuc07} or sine-Gordon model \cite{Cau03,Ahn04,Gou14,Rut20,San24} has also been investigated.

In several of these fermionic manifestations of the Casimir effect the boundaries break a fundamental symmetry: magnetic boundaries in a topological insulator break time-reversal symmetry, superconducting boundaries in a Josephson junction break ${\rm U}(1)$ charge conservation symmetry. Our objective here is to study the Casimir effect in the context of symmetric mass generation (SMG) \cite{Wan23,Wan19,Wan22,Ton22,Zen22,Zak26}, with boundaries that gap the fermions without symmetry breaking.

A boundary between an SMG-gapped region and a gapless region reflects the fermions in an anomalous way: In the simplest one-dimensional model (the so-called 3-4-5-0 model of Wang and Wen \cite{Wan23,Wan19}), two right-moving fermions carry charges 3 and 4, while two left-moving fermions have charges 5 and 0. At an interface between a gapped and an ungapped region, the reflection of an incident charge 3 or 4 into charge 5 or 0 must be accompanied by a many-body excitation to preserve the ${\rm U}(1)$ symmetry of charge conservation \cite{Boy20,vBe25,Ued26}.

Building on a recent bosonization analysis of the 3-4-5-0 model \cite{Zak26} (a representation by two decoupled sine-Gordon Hamiltonians \cite{Giamarchi} with scaling dimension $5K$, see Sec.\ \ref{sec_bosonized}), we study the imprint of this anomalous reflection on the Casimir force between two SMG-gapped boundaries, a distance $L$ apart. The six-fermion interaction of the 3-4-5-0 model gaps the system for Luttinger parameter $K\in(0,2/5)$. At $K=1/5$ (the Luther-Emery point of the sine-Gordon model \cite{Lut74,Col75,Cui26}) the 3-4-5-0 junction can be mapped onto a superconductor--normal-metal--superconductor (SNS) junction see Sec.\ \ref{sec_Andreevmapping}; The $L$-dependence of the Casimir force then follows from known results in a Josephson junction \cite{Bee23}.

Away from this special value of $K$ we obtain asymptotic results for $\hbar v/L$ much longer or much smaller than the gap $\Delta$, see Sec.\ \ref{sec_Casimir}. In the long-junction regime $\hbar v/L\ll \Delta$ the force decays $\propto 1/L^2$, while in the short-junction regime $\hbar v/L\gg \Delta$, and for $K\in(1/5,2/5)$, the force increases with decreasing $L$ as a power law $\propto L^{2-10K}$. For $K<1/5$ it tends to a finite value when the two gapped regions are brought into contact.

We relate the Casimir force to the ${\rm U}(1)$ charges of the 3-4-5-0 model in Sec.\ \ref{sec_charges}, representing it by the $(Q^2+Q'^2)/L^2$ repulsion of two integer-valued conserved ${\rm U}(1)$ charges $Q,Q'$, constrained by $3Q+4Q'=0\bmod 5$. We conclude in Sec.\ \ref{sec_conclude} by a comparison with results in the literature for the Casimir effect in the sine-Gordon and Luttinger models.

\section{Bosonized 3-4-5-0 junction}
\label{sec_bosonized}

\begin{figure}[tb]
\centerline{\includegraphics[width=0.9\linewidth]{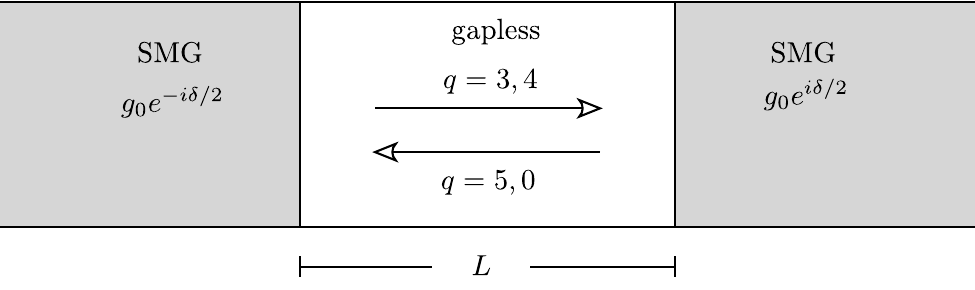}}
\caption{Schematic of a 3-4-5-0 junction: A gapless segment of length $L$ between two regions which are gapped by the SMG interaction of amplitude $g_0$ and phase difference $\delta$. The charge difference of the reflected chiral fermions is compensated by a many-body excitation, to preserve the ${\rm U}(1)$ charge symmetry. We compute the Casimir force between the two interfaces.}
\label{fig:layout}
\end{figure}

The 3-4-5-0 model \cite{Wan23,Wan19} of symmetric mass generation consists of a one-dimensional (1D) system with four species of chiral fermions: two right-movers with charges 3 and 4, and two left-movers with charges 5 and 0. The charges $q$ are chosen such that the system is  ``anomaly-free'', meaning that $\sum_q q^2$ is the same for right-movers and left-movers. This is required for the opening of an excitation gap without breaking ${\rm U}(1)$ symmetry \cite{Wan23,Wan19,Wan22,Ton22}. We assume that a gap is opened for $x<-L/2$ and $x>L/2$, while the interval $|x|<L/2$ remains gapless (see Fig.\ \ref{fig:layout}).

Upon bosonization of the chiral fermion operators $\Psi_q\sim e^{i\varphi_q}$, $q\in\{3,4,5,0\}$, the four charge-$q$ chiral bosonic fields $\varphi_q$ can be combined into two pairs of non-chiral fields, $(\phi_1,\theta_1)$ and $(\phi_2,\theta_2)$, defined by
\begin{equation}
\begin{split}
&2 \sqrt{5}\,\phi_1=(\varphi_3-2 \varphi_4)+(\varphi_5+2 \varphi_0),\\
&2 \sqrt{5}\,\phi_2=(2 \varphi_3+\varphi_4)-(2 \varphi_5-\varphi_0),\\
&2 \sqrt{5}\,\theta_1=(\varphi_3-2 \varphi_4)-(\varphi_5+2 \varphi_0),\\
&2 \sqrt{5}\,\theta_2=(2 \varphi_3+\varphi_4)+(2 \varphi_5-\varphi_0).
\end{split}\label{phivarphi}
\end{equation}
(The parentheses distinguish left-movers from right-movers.) This orthogonal transformation separates the neutral $\phi_{j}$ fields from the charged $\theta_{j}$ fields.\footnote{The fields $\phi_{j}$ are neutral because invariant under the U(1) transformation $\varphi_q\mapsto q\chi+\varphi_q$} It is useful because the gapping interaction couples only to the neutral fields, in order to preserve the U(1) symmetry.

As shown in Ref.\ \onlinecite{Zak26}, the bosonized 3-4-5-0 Hamiltonian takes the form of two \textit{decoupled} Luttinger liquids with a cosine potential (sine-Gordon model \cite{Giamarchi}),
\begin{subequations}
\label{eq:HLL}
\begin{align}
H_{3450}={}&\sum_{j=1}^2\int dx \,{\cal H}_{j}(x),\\
 {\cal H}_{j}(x)={}&\frac{\hbar v_j}{2\pi}
 \left[K_{j}(\partial_x\theta_{j})^2+\frac{1}{K_{j}}
       (\partial_x\phi_{j})^2\right]\nonumber\\
       &- g_{j}(x) \cos\bigl[2\sqrt5\,\phi_{j}-\alpha_{j}(x)\bigr].
\end{align}
\end{subequations}
The Luttinger parameter $K_{j}\in(0,1)$ is reduced below unity by a repulsive Hubbard interaction $U_{j}\propto(\partial_x\phi_{j})^2-(\partial_x\theta_{j})^2$ (with a positive coefficient). This density-density interaction renormalizes the velocity $v_j=2v_0K_j/(1+K_j^2)$, it mixes left-movers and right-movers, but it does not open a gap. We take the Hubbard interaction uniform, so that $K_j$ and $v_j$ do not depend on $x$; only the gapping interaction is switched off in the junction.

The gapping cosine potential has a complex interaction parameter of amplitude $g_{j}>0$ and phase $\alpha_{j}\in(0,2\pi)$. We choose the profile
\begin{equation}
g_j(x)e^{i\alpha_j(x)}=\begin{cases}
g_0 e^{-i\delta_j/2}&\text{for}\;\;x<-L/2,\\
g_0 e^{i\delta_j/2}&\text{for}\;\;x>L/2,\\
0&\text{for}\;\; -L/2<x<L/2.
\end{cases}\label{gjprofile}
\end{equation}
An excitation gap opens up for $|x|>L/2$ if $K_{1},K_2<2/5$ \cite{Zak26}. We will now show that for $K_1=K_2=1/5$ the system can be mapped onto an SNS Josephson junction.

\section{Mapping of the 3-4-5-0 junction at \emph{K=1/5} onto a Josephson junction}
\label{sec_Andreevmapping}

\subsection{Refermionization}

We first rewrite the bosonized 3-4-5-0 Hamiltonian \eqref{eq:HLL} in terms of right-moving (R) or left-moving (L) bosonic fields, accounting for the mixing of the bare fields by the Hubbard interaction,
\begin{equation}
\begin{split}
&\eta_{{\rm R}j}=\sqrt{K_j}\,\theta_j+ \frac{1}{\sqrt{K_j}}\phi_j,\\
&\eta_{{\rm L}j}=\sqrt{K_j}\,\theta_j- \frac{1}{\sqrt{K_j}}\phi_j.
\end{split}
\end{equation}
This gives
\begin{subequations}
\label{eq:HLL2}
\begin{align}
&H_{3450}=\sum_{j=1}^2\int dx \,{\cal H}_{j}(x),\\
&{\cal H}_{j}(x)=\frac{\hbar v_j}{4\pi}
 \left[(\partial_x\eta_{{\rm R}j})^2+
       (\partial_x\eta_{{\rm L}j})^2\right]\nonumber\\
       &\qquad- g_{j}(x) \cos\bigl[\sqrt{5K_j}\,(\eta_{{\rm R}j}-\eta_{{\rm L}j})-\alpha_{j}(x)\bigr].
\end{align}
\end{subequations}

The choice $K_j=1/5$ is the Luther-Emery point of the sine-Gordon model \cite{Lut74,Col75,Cui26}, where Eq.\ \eqref{eq:HLL2} is the bosonized Hamiltonian of a fermionic spinor\footnote{The relation \eqref{refermionization} between the fermionic and bosonic fields assumes zero chemical potential (half-filled band on a lattice) \cite{Giamarchi}.}
\begin{equation}
\Psi_j=\begin{pmatrix}
\psi_{{\rm R}j}\\
\psi_{{\rm L}j}
\end{pmatrix}=\frac{1}{\sqrt{2\pi a_0}}\begin{pmatrix}
e^{i\eta_{{\rm R}j}}\\
e^{i\eta_{{\rm L}j}}
\end{pmatrix},\label{refermionization}
\end{equation}
gapped by a complex Dirac mass
\begin{equation}
\Delta_j(x)=-\pi a_0 g_j(x)e^{i\alpha_j(x)}.
\end{equation}
Here $a_0$ is a short-distance cutoff length.\footnote{On a lattice the cutoff length $a_0$ is set by the lattice constant, up to a non-universal numerical factor of order unity.} The fermionic Hamiltonian $H_{3450}$ at this special point $K_1=K_2=1/5$ is
\begin{align}
H_{3450}\bigr|_{K=1/5}=\sum_{j=1}^2 \int dx\,\Psi_j^\dagger
\begin{pmatrix}
 v_{\rm F} p_x&\Delta_j(x)\\
\Delta_j^\ast(x)&- v_{\rm F} p_x
\end{pmatrix}
\Psi_j^{\vphantom{\dagger}},\label{H15def}
\end{align}
with $v_1=v_2=5v_0/13\equiv v_{\rm F}$.

\subsection{Mapping onto the BCS Hamiltonian}

To make contact with the superconducting problem we consider the mean-field BCS Hamiltonian, for 1D spin-$\sigma$ fermionic fields $\psi_\sigma(x)$,
\begin{align}
H_{\rm BCS}={}&\int dx\,\Bigl(\sum_{\sigma=\uparrow,\downarrow}\psi_\sigma^\dagger (p^2/2m-E_{\rm F})\psi_\sigma\nonumber\\
&-\Delta(x)\psi_\uparrow^\dagger\psi_\downarrow^\dagger-\Delta^*(x)\psi_\downarrow\psi_\uparrow\Bigr).
\end{align}
Assuming $|\Delta|\ll E_{\rm F}$, the relevant regime in a metal, we expand around the Fermi wave vector: $\psi_\sigma=e^{ik_{\rm F}x}\psi_{{\rm R}\sigma}+e^{-ik_{\rm F}x}\psi_{{\rm L}\sigma}$, with right-moving electrons near $k_{\rm F}$ and left-movers near $-k_{\rm F}$. 

Dropping the terms that oscillate as $e^{\pm2ik_{\rm F}x}$, the Hamiltonian separates into two parts,
\begin{subequations}
\label{eq:HA}
\begin{align}
H_{\rm BCS}={}&H_A[\Delta]+H_B[\Delta],\\
H_A[\Delta]={}&\int dx\,\Bigl(\psi_{{\rm R}\uparrow}^\dagger v_{\rm F}p_x\psi_{{\rm R}\uparrow}^{\vphantom{\dagger}}-\psi_{{\rm L}\downarrow}^\dagger v_{\rm F}p_x\psi_{{\rm L}\downarrow}^{\vphantom{\dagger}}\nonumber\\
&-\Delta\,\psi_{{\rm R}\uparrow}^\dagger\psi_{{\rm L}\downarrow}^\dagger-\Delta^*\psi_{{\rm L}\downarrow}\psi_{{\rm R}\uparrow}\Bigr),\\
H_B[\Delta]={}&\int dx\,\Bigl(\psi_{{\rm R}\downarrow}^\dagger v_{\rm F}p_x\psi_{{\rm R}\downarrow}^{\vphantom{\dagger}}-\psi_{{\rm L}\uparrow}^\dagger v_{\rm F}p_x\psi_{{\rm L}\uparrow}^{\vphantom{\dagger}}\nonumber\\
&+\Delta\,\psi_{{\rm R}\downarrow}^\dagger\psi_{{\rm L}\uparrow}^\dagger+\Delta^*\psi_{{\rm L}\uparrow}\psi_{{\rm R}\downarrow}\Bigr).
\end{align}
\end{subequations}
The two parts have the same spectrum, because $H_B[\Delta]$ equals $H_A[-\Delta]$ upon interchange of the spin labels, and the sign of $\Delta$ can be removed by a gauge transformation.

We compare with the Hamiltonian \eqref{H15def}, written out in components,
\begin{align}
H_{3450}\bigr|_{K=1/5}{}&=\sum_{j=1}^2\int dx\,\Bigl[\psi_{{\rm R}j}^\dagger v_{\rm F}p_x\psi_{{\rm R}j}^{\vphantom{\dagger}}-\psi_{{\rm L}j}^\dagger v_{\rm F}p_x\psi_{{\rm L}j}^{\vphantom{\dagger}}\nonumber\\
&+\Delta_j(x)\psi_{{\rm R}j}^\dagger\psi_{{\rm L}j}^{\vphantom{\dagger}}+\Delta_j^\ast(x)\psi_{{\rm L}j}^\dagger\psi_{{\rm R}j}^{\vphantom{\dagger}}\Bigr].
\end{align}
Upon identification of $\psi_{{\rm R}j}\leftrightarrow\psi_{{\rm R}\uparrow}$ and $\psi_{{\rm L}j}\leftrightarrow\psi_{{\rm L}\downarrow}^\dagger$, the Dirac mass term turns into the pairing term while the kinetic term is left unchanged up to a constant. This gives the following Hamiltonian correspondence in the refermionized description:
\begin{equation}
H_{3450}\big|_{K=1/5}\cong H_A[-\Delta_1]+H_A[-\Delta_2].\label{operatoridentity}
\end{equation}
(A uniform sign change of the pair potential is a uniform shift of its phase by $\pi$, without effect on the spectrum.) 

The fermion fields $\Psi_j$ are solitons of the 3-4-5-0 model. The conserved soliton number $N_j=N_{{\rm R}j}+N_{{\rm L}j}$ maps onto $N_{{\rm R}\uparrow}-N_{{\rm L}\downarrow}$, the spin of sector $A$, while the chiral number $J_j=N_{{\rm R}j}-N_{{\rm L}j}$ of $\Psi_j$ maps onto the charge $N_{{\rm R}\uparrow}+N_{{\rm L}\downarrow}$ of the superconducting problem, conjugate to the phase $\alpha_j$, which is not conserved. The conserved charges of the 3-4-5-0 model thus play the role of the spin of the Josephson junction.

\subsection{Ground state energy}

To relate the ground state energy of the 3-4-5-0 Hamiltonian to that of the BCS Hamiltonian, we still need to establish that the state space of the original four fermions $\psi_q$, with integer particle numbers $n_3,n_4,n_5,n_0$, becomes the unrestricted state space of two independent Dirac fermions $\Psi_j$, with integer soliton numbers $N_1,N_2$ and chiral numbers $J_1,J_2$. The state-space correspondence is not automatic because the solitons are nonlocal in terms of the original chiral fermions. We work this out in App.\ \ref{app:sectors}, with the following results.

The state-space correspondence is complete for the soliton numbers. Since
\begin{equation}
\begin{split}
&N_1=n_3-2n_4-n_5-2n_0,\\
&N_2=2n_3+n_4+2n_5-n_0,
\end{split}
\end{equation}
any integer choice for the $n_q$'s gives integer $N_j$ and since
\begin{align}
&(n_3,n_4,n_5,n_0)=(0,1,-1,-1)\rightarrow (N_1,N_2)=(1,0),\nonumber\\
&(n_3,n_4,n_5,n_0)=(1,1,-1,0)
\rightarrow (N_1,N_2)=(0,1),
\end{align}
every integer pair $(N_1,N_2)$ can be reached by integer combinations of the $n_q$'s. Also note that the parity of $N_1+N_2$ is the fermion parity of the 3-4-5-0 model.

The correspondence is incomplete for the chiral numbers. For the four chiral fermions one has
\begin{equation}
\begin{split}
J_1&=\tfrac{1}{5}(n_3-2n_4+n_5+2n_0)\in N_1-\tfrac{2}{5}N_2+2\mathbb{Z},\\
J_2&=\tfrac{1}{5}(2n_3+n_4-2n_5+n_0)\in N_2+\tfrac{2}{5}N_1+2\mathbb{Z},
\end{split}
\end{equation}
while for the Dirac fermions $J_j\in N_j+2\mathbb{Z}$. The sets only coincide if $N_1$ and $N_2$ are both integer multiples of 5 (even or odd multiples dependent on the parity of $J_j$). In particular, in the ground-state sector $N_1=N_2=0$ the two state spaces coincide.

We conclude that the spinless 3-4-5-0 junction at $K_1=K_2=1/5$ is equivalent, as far as its ground state energy is concerned, to two copies of sector $A$ of the Josephson junction, each with its own pair potential.

We apply the mapping to the ground state energy ${\cal E}_{\rm SNS}(\delta)$ of the SNS junction with pair potential profile
\begin{equation}
\Delta(x)=\begin{cases}
\Delta_0 e^{-i\delta/2}&\text{for}\;\;x<-L/2,\\
\Delta_0 e^{i\delta/2}&\text{for}\;\;x>L/2,\\
0&\text{for}\;\; -L/2<x<L/2.
\end{cases}\label{Deltaprofile}
\end{equation}
Sectors $A$ and $B$ have the same spectra and thus contribute equally to ${\cal E}_{\rm SNS}$. In view of Eq.\ \eqref{operatoridentity}, the ground state energy of the 3-4-5-0 junction with gapping profile \eqref{gjprofile}, for $g_0\equiv \Delta_0/\pi a_0$, is then given by
\begin{equation}
{\cal E}_{3450}\bigl|_{K=1/5}=\tfrac{1}{2}\sum_{j=1}^2{\cal E}_{\rm SNS}(\delta_j).\label{E3450ESNS}
\end{equation}

\section{Casimir force}
\label{sec_Casimir}

The Casimir energy ${\cal E}_{\rm C}$ of the junction is the $L$-dependent contribution to the ground state energy, excluding contributions that do not depend on the separation of the interfaces. We assume that the gapped intervals $(-D-L/2,-L/2)\cup(L/2,D+L/2)$ extend over a fixed $L$-independent length $D$, to exclude a spurious contribution $\propto L$ to the ground state energy. For penetration length $\xi=\hbar v/\Delta$ small compared to the gapped interval length $D$, the Casimir energy does not depend on $D$ and we may use the profile \eqref{gjprofile} with $D\rightarrow\infty$. The derivative ${\cal F}_{\rm C}=-d{\cal E}_{\rm C}/dL$ is the force that one interface exerts on the other. A negative ${\cal F}_{\rm C}$ favors decreasing the interface separation (attractive Casimir force).

\subsection{Full length dependence at \emph{K=1/5}}

\begin{figure}[tb]
\centerline{\includegraphics[width=0.9\linewidth]{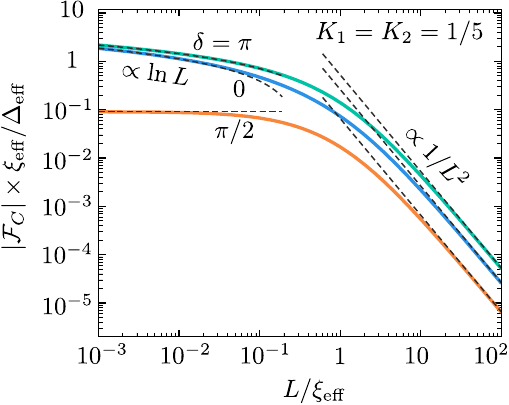}}
\caption{Length dependence of the Casimir force at $K_1=K_2=1/5$, from Eq.\ \eqref{eq:LE}, for three values of the phase difference $\delta_1=\delta_2\equiv\delta$ (solid curves), with the long-junction and short-junction asymptotes \eqref{eq:long} and \eqref{eq:short} (dashed). The force is attractive for $\delta=0,\pi/2$ and repulsive for $\delta=\pi$.}
\label{fig:scaling}
\end{figure}

The Casimir energy of the spin-degenerate SNS junction, with pair potential profile \eqref{Deltaprofile}, is known \cite{Bee23}. The correspondence \eqref{E3450ESNS} then gives the Casimir energy of the 3-4-5-0 junction, at $K_1=K_2=1/5$ and with the identifications $v_1=v_2\equiv v_{\rm F}$, $g_0\equiv \Delta_0/\pi a_0$:
\begin{equation}
\begin{split}
&{\cal E}_{\rm C}=-\frac{\Delta_0}{\pi}\sum_{j=1}^2\operatorname{Re}\int_0^\infty d\omega\,\ln\Bigl[1+\beta(\omega)^2e^{i\delta_j-2\omega L/\xi}\Bigr],\\
&\beta(\omega)=\omega-\sqrt{1+\omega^2},\;\;\xi=\frac{\hbar v_{\rm F}}{\Delta_0}.
\end{split}
\label{eq:LE}
\end{equation}
The small and large-$L$ asymptotics of the force are\footnote{The ${\cal O}(1)$ term in Eq.\ \eqref{eq:short} has a closed form expression: $-\gamma_{\rm E}\cos\delta_j-\tfrac{1}{2}+\tfrac{1}{2}\delta_j\sin\delta_j$, $|\delta_j|<\pi$, with $\gamma_{\rm E}$ the Euler constant, which we use in Fig.\ \ref{fig:scaling}.}
\begin{align}
L\gg\xi:{\cal F}_{\rm C}={}&\frac{\pi\hbar v_{\rm F}}{L^2}\sum_{j=1}^2\Bigl(\frac{[\arccos(\cos\delta_j)]^2}{8\pi^2}\nonumber\\
&-\frac1{24}+{\cal O}(\xi/L)\Bigr),\label{eq:long}\\
L\ll\xi: {\cal F}_{\rm C}={}&-\frac{\Delta_0}{2\pi\xi}\sum_{j=1}^2\Bigl(\ln(\xi/L)\cos\delta_j+{\cal O}(1)\Bigr).\label{eq:short}
\end{align}
These results are plotted in Fig.\ \ref{fig:scaling}.

\subsection{Long junction for any \emph{K\ \textless\ 2/5}}

In the large-$L$ regime, for energies $\hbar v/L$ small compared to $\Delta$, we may treat the gapped regions as a boundary condition at $x=\pm L/2\equiv x_\pm$, neglecting the finite penetration length. We may then find the Casimir energy for any value of the Luttinger parameters in the gapped range $K_1,K_2<2/5$.

Since the gapped $\phi_{j}$ fields in the 3-4-5-0 junction Hamiltonian \eqref{eq:HLL} are pinned to the minimum of the cosine potential, at $\cos\bigl(2\sqrt5\,\phi_{j}-\alpha_{j}\bigr)=1$, we have
\begin{equation}
\quad 2\sqrt5\,\phi_{j}(x_\pm)=\alpha^\pm_{j}+2\pi n^\pm_{j},
 \;\; n^\pm_{j}\in\mathbb{Z}.
 \label{eq:pinning}
\end{equation}
The difference at the two end points is
\begin{equation}
\begin{split}
& \phi_{j}(L/2)-\phi_{j}(-L/2)
   =\frac{\delta_{j}+2\pi w_{j}}{2\sqrt5},\;\; w_{j}\in\mathbb{Z},\\
&\delta_{j}=\alpha_{j}^{+}-\alpha_{j}^{-},\;\;w_{j}=n_{j}^{+}-n_{j}^{-}.
\end{split}
 \label{eq:winding}
\end{equation}

The contribution $E_\phi$ of the $\phi$ fields to the energy \eqref{eq:HLL}, at fixed end points, is minimized by a linear profile,
\begin{equation}
 \phi_{j}(x)=\phi_{j}(-L/2)+\frac{x+L/2}{L}
       \frac{\delta_{j}+2\pi w_{j}}{2\sqrt5},
\end{equation}
resulting in the energy
\begin{equation}
E_\phi=
\frac{\pi \hbar }{10L}\sum_{j=1}^2\frac{v_j}{K_{j}}
      (w_{j}+\delta_{j}/2\pi)^2.
        \label{eq:interacting}
\end{equation}
This is the energy produced by the winding $\delta_{j}+2\pi w_{j}$ of the phase of the pinning potential between the two gapped regions. The winding represents a topological defect that appears in order to ensure charge conservation when electrons with charge 0 or 5 are reflected into electrons with charge 3 or 4 \cite{vBe25}.

For the ground-state energy we minimize over the winding numbers and add the contribution $-\tfrac{1}{24}\pi\hbar (v_1+v_2)/L$ from vacuum fluctuations of the fields. The resulting force is
\begin{equation}
{\cal F}_{\rm C}= \frac{\pi \hbar }{10L^2}\sum_{j=1}^2\frac{v_j}{K_{j}}\min_{w_{j}\in\mathbb{Z}}(w_{j}+\delta_{j}/2\pi)^2-\frac{\pi\hbar}{24 L^2}(v_1+v_2).\label{FClongallK}
\end{equation}
For $K_1=K_2=1/5$, $v_1=v_2\equiv v_{\rm F}$ this is Eq.\ \eqref{eq:long}.

\subsection{Short junction for \emph{1/5\ \textless\ K\ \textless\ 2/5}}

In the opposite small-$L$ regime the cosine potential is a weak perturbation at the relevant energies $\hbar v/L\gg\Delta$. The continuum description requires that $L$ remains large compared to the short-distance cutoff $a_0$, so the short-junction regime is the window $a_0\ll L\ll\hbar v/\Delta$. 

Second-order perturbation theory gives the leading $L$-dependence of the force (see App.\ \ref{app:pert}) for $K_1,K_2\in(1/5,2/5)$,
\begin{equation}
{\cal F}_{\rm C}=-\sum_{j=1}^2 \frac{\sqrt\pi(g_0a_0)^2\Gamma(5K_j-\frac{1}{2})}{4\hbar v_j(5K_j-1)\Gamma(5K_j)}\left(\frac{a_0}{L}\right)^{10K_j-2}\cos\delta_j .
\label{eq:Fpert}
\end{equation}
For $K_1=K_2\equiv K\in(1/5,2/5)$ this can be written more informatively as
\begin{equation}
{\cal F}_{\rm C}=-C(K) \frac{\Delta_{\rm eff}}{\xi_{\rm eff}}\left(\frac{\xi_{\rm eff}}{L}\right)^{10K-2}(\cos\delta_1+\cos\delta_2),\label{FCresultshort}
\end{equation}
with $\Delta_{\rm eff}\propto g_0^{1/(2-d)}$ the soliton mass of the sine-Gordon model at scaling dimension $d=5K$, and $\xi_{\rm eff}=\hbar v/\Delta_{\rm eff}$ the corresponding length scale. (For $K>1/5$ the soliton is the lowest excitation \cite{Giamarchi}, so then $\Delta_{\rm eff}$ is the excitation gap and $\xi_{\rm eff}$ the penetration length into the gapped region.)

The numerical coefficient $C(K)$ follows from the known relation between $g_0$ and $\Delta_{\rm eff}$ in the sine-Gordon model \cite{Zam95},
\begin{align}
C(K)={}&\frac{\Gamma\bigl(d-1/2\bigr)\,\Gamma\bigl(d/2\bigr)^{2}}
{\pi^{3/2}\,(d-1)\,\Gamma(d)\,\Gamma\bigl(1-d/2\bigr)^{2}}\nonumber\\
&\times\left[\frac{\sqrt{\pi}\,\Gamma\bigl(\tfrac{1+\zeta}{2}\bigr)}{2\,\Gamma\bigl(\tfrac{\zeta}{2}\bigr)}\right]^{4-2d},\;\;\zeta=\frac{d}{2-d}.\label{CKdef}
\end{align}
The power law increase of the force with decreasing $L$ applies to generic phase differences, such that the coefficient $\cos\delta_1+\cos\delta_2$ in Eq.\ \eqref{FCresultshort} does not vanish. On a lattice the increase stops when $L$ becomes of the order of the lattice constant.

The logarithmic $L$-dependence \eqref{eq:short} in the limit $K\rightarrow 1/5$ is the marginal case between a force that stays finite for $L\to0$ ($K<1/5$) and one that grows as a power ($1/5<K<2/5$). As we show next, the constant $L\rightarrow 0$ limit of the force for $K<1/5$ and $\delta=0$ can be obtained directly from the known bulk energy density of the sine-Gordon model \cite{Des91,Zam95}, without having to solve the junction problem.

\subsection{Short junction for \emph{K\ \textless\ 1/5}}

The Hamiltonian \eqref{eq:HLL} depends on $L$ only through the position of the interfaces. A rigid displacement $\delta x$ of a gapped region of width $D$ removes the gapping interaction at its inner interface, at $|x|=L/2$, and adds it at its outer interface, at $|x|=L/2+D$. By the Hellmann-Feynman theorem,
\begin{equation}
{\cal F}_{\rm C}=-\frac{d}{dL}{\cal E}_{\rm C}=-\bigl\langle\frac{d}{dL}H_{3450}\bigr\rangle,
\end{equation}
the force is then the difference of the expectation values at the outer and inner interfaces of the local gapping interaction ${\cal V}_j(x)=\cos[2\sqrt5\,\phi_j(x)-\alpha_j(x)]$,
\begin{align}
{\cal F}_{\rm C}={}&\tfrac{1}{2}g_0\sum_{j=1}^2\Bigl(\bigl\langle{\cal V}_j(L/2+D)\bigr\rangle-\bigl\langle{\cal V}_j(L/2)\bigr\rangle\nonumber\\
&+\langle{\cal V}_j(-L/2-D)\bigr\rangle-\bigl\langle{\cal V}_j(-L/2)\bigr\rangle\Bigr).\label{HFidentity1}
\end{align}

Each of the four interfaces contributes with a factor $1/2$, because it is displaced by $\delta x=\delta L/2$. The interfaces at $x$ and at $-x$ contribute equally: for the profile \eqref{gjprofile} the Hamiltonian is invariant under $x\mapsto-x$, $\phi_j\mapsto-\phi_j$, hence $\langle{\cal V}_j(x)\rangle=\langle{\cal V}_j(-x)\rangle$. We may thus reduce the number of terms,
\begin{equation}
{\cal F}_{\rm C}=g_0\sum_{j=1}^2\Bigl(\bigl\langle{\cal V}_j(L/2+D)\bigr\rangle-\bigl\langle{\cal V}_j(L/2)\bigr\rangle\Bigr).\label{HFidentity}
\end{equation}

The outer interface faces a semi-infinite gapless region. A shift of such an isolated interface converts gapped into gapless material, at an energy cost $-e_j$ per unit length, with $e_j<0$ the energy density of the gapped phase relative to the gapless phase. Hence $g_0\langle{\cal V}_j(L/2+D)\rangle=-e_j$, independent of $\delta_j$ (the phase at an isolated interface can be removed by a shift of $\phi_j$). 

So far we made no assumptions on $L$ or $\delta_j$; we now take the $L\rightarrow 0$ limit at zero phase difference, $\delta_j=0$. In this limit the two gapped regions merge into a uniform one, and the expectation value for the energy density at the inner interface is the bulk value, 
\begin{equation}
g_0\langle{\cal V}_j\rangle_{\rm bulk}=-g_0\bigl\langle \frac{d}{dg_0}{\cal H}_j\bigr\rangle_{\rm bulk}=-g_0\frac{de_j}{dg_0}=-\frac{2e_j}{2-5K_j}.
\end{equation}
We again used the Hellmann-Feynman theorem, and the scaling $e_j\propto g_0^{2/(2-5K_j)}$.

Combining the results from inner and outer surfaces, we have the force
\begin{align}
\lim_{L\rightarrow0}{\cal F}_{\rm C}\bigr|_{\delta_j=0}={}&\sum_{j=1}^2\bigl(-e_j+\frac{2e_j}{2-5K_j}\bigr)\nonumber\\
={}&\sum_{j=1}^2\zeta_je_j,\;\;\zeta_j=\frac{5K_j}{2-5K_j}.
\end{align}
The bulk energy density of the sine-Gordon model is known \cite{Des91,Zam95},
\begin{equation}
e_j=-\frac{\Delta_{{\rm eff},j}^2}{4\hbar v_j}\tan\frac{\pi\zeta_j}{2},\label{eSG}
\end{equation}
with $\Delta_{{\rm eff},j}$ the soliton mass. We thus arrive at the contact force
\begin{equation}
\lim_{L\rightarrow0}{\cal F}_{\rm C}\bigr|_{\delta_j=0}=-\sum_{j=1}^2\frac{\Delta_{{\rm eff},j}^2}{4\hbar v_j}\,\zeta_j\tan\frac{\pi\zeta_j}{2}.\label{FCzero}
\end{equation}
The force is attractive. It diverges as $(1-5K_j)^{-1}$ for $K_j\rightarrow1/5$, with the coefficient that follows from Eq.\ \eqref{eq:Fpert}. For $K_j\rightarrow0$ at fixed $\Delta_{{\rm eff},j}$ it vanishes as $K_j^2$.

In terms of the coupling strength $g_0$, upon elimination of $\Delta_{{\rm eff},j}$ by means of the relation of Ref.\ \onlinecite{Zam95}, Eq.\ \eqref{FCzero} reads
\begin{subequations}
\label{FCshortresult}
\begin{align}
&\lim_{L\rightarrow0}{\cal F}_{\rm C}\bigr|_{\delta_j=0}=-\sum_{j=1}^2 A(K_j)\,\frac{\hbar v_j}{a_0^2}
\left(\frac{g_0a_0^2}{\hbar v_j}\right)^{1+\zeta_j},\\
&A(K)=\frac{\zeta\tan(\pi\zeta/2)\,\Gamma(\zeta/2)^2}{\pi\,\Gamma\bigl(\tfrac{1+\zeta}{2}\bigr)^2}
\left[\frac{\pi\,\Gamma(1-5K/2)}{2\,\Gamma(5K/2)}\right]^{1+\zeta}.
\end{align}
\end{subequations}

\section{Junction charges}
\label{sec_charges}

The winding energy \eqref{eq:interacting} in the long-junction regime can be written equivalently in term of two ${\rm U}(1)$ charges $Q,Q'$ of the junction, defined in terms of the $\phi$ fields \eqref{phivarphi} by the charge densities
\begin{equation}
\begin{split}
&\rho_Q=\frac{\sqrt{5}}{\pi}\frac{d}{dx}(2\phi_2-\phi_1),\\
&\rho_{Q'}=\frac{\sqrt{5}}{\pi}\frac{d}{dx}(\phi_2+2\phi_1).
\end{split}
\end{equation}
Both $Q$ and $Q'$ are conserved in the 3-4-5-0 model: $Q$ is the electrical charge of the fermions, $Q'$ is a non-electrical ``flavour'' charge \cite{Wan23,Wan19}.

For $K_1=K_2\equiv K$, $v_1=v_2\equiv v$, the winding energy becomes
\begin{align}
&E_\phi=E_{\rm W}(Q^2+Q'^2),\;\;E_{\rm W}=\frac{\pi\hbar v}{50LK},\label{EphiQQ}\\
\begin{split}
&Q=-(w_1+\delta_1/2\pi)+2(w_2+\delta_2/2\pi),\\
&Q'=2(w_1+\delta_1/2\pi)+(w_2+\delta_2/2\pi).
\end{split}\label{eq:charge}
\end{align}
The fact that $w_1,w_2$ are integers constrains the allowed values of $Q,Q'$. At zero phase difference ($\delta_1=0=\delta_2$) the allowed charges form the lattice
\begin{equation}
 \Lambda=\{(Q,Q')\in\mathbb{Z}^2:3Q+4Q'=0\pmod5\}.
 \label{eq:lattice}
\end{equation}
At nonzero phase difference the lattice is shifted by a fractional offset.

\begin{figure}[tb]
\centerline{\includegraphics[width=0.9\linewidth]{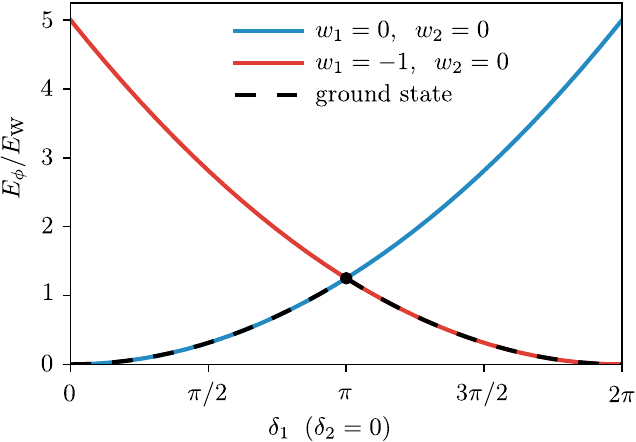}}
\caption{Plot of $E_\phi/E_W$ as a function of $\delta_1$ for $\delta_2=0$, from Eq.\ \eqref{EphiQQ}. The two branches are distinguished by the charges $(Q,Q')$; at the crossing point one branch has $(Q,Q')=(-\tfrac{1}{2},1)$ the other branch $(\tfrac{1}{2},-1)$, so a charge increment $(1,-2)$.
}
\label{fig:winding}
\end{figure}

The phase-dependent part of the Casimir energy can therefore be interpreted as a charging energy: the walls of a charged junction repel each other, with a $1/L^2$ force whose coefficient, at zero phase difference, takes on discrete values on the lattice $\Lambda$. (At nonzero phase difference the winding numbers remain integer, but the energy varies continuously with $\delta_j$.) At $\delta_1=0=\delta_2$ the lowest value of Eq.\ \eqref{EphiQQ} at a given electrical charge $Q\in\{0,1,2,3,4,5,\dots\}$, minimized over the allowed values of $Q'$, is $E_\phi/E_{\rm W}\in\{0,5,5,10,20,25,\dots\}$; the degeneracy at $Q=1$ and $Q=2$ corresponds to charges $(Q,Q')=(1,-2)$ and $(2,1)$.

The minimization ${\rm min}_{w_j}$ in Eq.\ \eqref{FClongallK} assumes that the junction can exchange charge with a reservoir, in units $(Q,Q')=(1,-2)$ (a change of $w_1$) and $(2,1)$ (a change of $w_2$). The ground state then switches from $w_j=0$ to $w_j=-1$ at $\delta_j=\pi$, where the force has a cusp (see Fig.\ \ref{fig:winding}). The two states that cross at $\delta_1=\pi$, $\delta_2=0$ have $(Q,Q')=(-\tfrac12,1)$ and $(\tfrac12,-1)$. They differ by one unit of the electrical charge $Q$, so the crossing is protected by ${\rm U}(1)$ charge conservation. 

Without a reservoir, the junction remains on a branch of fixed winding number when the phase $\alpha_j^+$ of the right gapped region is varied slowly on the scale of the gap. The density $n_j=(\sqrt5/\pi)\partial_x\phi_j$, with $\rho_Q=2n_2-n_1$ and $\rho_{Q'}=2n_1+n_2$, satisfies the continuity equation $\partial_t n_j+\partial_x I_j=0$ with current $I_j=-(\sqrt5/\pi)\partial_t\phi_j$. Since the field in the gapped region is locked to the phase, by Eq.\ \eqref{eq:pinning}, a current $I_j=-(2\pi)^{-1}d\alpha_j^+/dt$ flows uniformly through that region, towards the junction. Advancing $\alpha_j^+$ by $2\pi$ thus pumps the charge $(-1,2)$ (for $j=1$) or $(2,1)$ (for $j=2$) through the right gapped region into the junction, from the far end of the system --- a Thouless pump driven by the interaction phase $\delta_j$.

This phase difference can be interpreted as a gate voltage on the junction. The shift $\phi_j\mapsto\phi_j+\alpha_j(x)/2\sqrt5$, with $\alpha_j(x)$ a smooth interpolation in the junction between $-\delta_j/2$ and $\delta_j/2$, removes the phase from the cosine in Eq.\ \eqref{eq:HLL}, at the expense of a potential $U_j$ that couples to the density $n_j=(\sqrt5/\pi)\partial_x\phi_j$,
\begin{equation}
\begin{split}
&{\cal H}_j\mapsto{\cal H}_j\bigr|_{\alpha_j=0}+U_j n_j+\frac{5K_j}{2\pi\hbar v_j}U_j^2,\\
&U_j=\frac{\hbar v_j}{10K_j}\frac{d\alpha_j}{dx}.
\end{split}
\end{equation}

The shift also changes the density, $n_j\mapsto n_j+(2\pi)^{-1}d\alpha_j/dx$, so that the soliton number of the junction with the potential $U_j$ is the integer $w_j$, while the offset $\delta_j/2\pi$ in Eq.\ \eqref{eq:charge} appears as a background charge. Since $\rho_Q=2n_2-n_1$ and $\rho_{Q'}=2n_1+n_2$, this is a combination of an electrostatic potential and a flavour potential in the junction, related to the phase difference by
\begin{equation}
\delta_j=\frac{10K_j}{\hbar v_j}\int_{-L/2}^{L/2}U_j(x)\,dx.
\end{equation}
The winding energy \eqref{EphiQQ} is the analogue of the charging energy of a Coulomb island, with $\delta_j/2\pi$ in the role of the gate charge.

\section{Conclusion}
\label{sec_conclude}

In summary, we have calculated the Casimir force between two interfaces, a length $L$ apart, that confine chiral one-dimensional fermions by symmetric mass generation, in the framework of the 3-4-5-0 model. We have obtained the force asymptotically in the long-junction and short-junction regimes, and for any junction length at Luttinger parameter $K=1/5$ by a mapping of the 3-4-5-0 junction onto a Josephson junction.

These specific results have parallels in other one-dimensional strongly interacting systems, and it is instructive to make the connection.

The long-junction force is an instance of the finite-size scaling of a conformal field theory on a strip \cite{Giamarchi}: The energy is $(\pi\hbar v/L)(h-c/24)$, with $c$ the central charge and $h$ the lowest conformal weight allowed by the two boundary conditions \cite{Blo86,Aff86,Car86}. Here each of the two bosonic fields has $c=1$ and $h_j=(w_j+\delta_j/2\pi)^2/10K_j$, with winding number $w_j\in\mathbb{Z}$. What is special for the 3-4-5-0 model is the representation of $h$ as the charging energy of two conserved ${\rm U}(1)$ charges on the lattice $3Q+4Q'=0\bmod 5$, with a fractional offset that is set by the phase of the interaction.

The short-junction power law is analogous to powerlaws for an impurity in a Luttinger liquid \cite{Kan92}, the Josephson current through a Luttinger liquid \cite{Faz95,Mas96,Aff00}, and the Casimir interaction between two impurities \cite{Rec05,Fuc07}. For two point-like perturbations of scaling dimension $d$ the interaction energy varies as $L^{1-2d}$; for two half-lines, as is the case here, it varies as $L^{3-2d}$, hence the force as $L^{2-2d}$. What is special for the 3-4-5-0 model is that the dimension $d=5K$ can be tuned through the free-fermion (Luther-Emery) point $d=1$ by the Hubbard interaction.

The crossover between the two regimes is governed by a single function of $L/\xi$. Such scaling functions have been calculated for integrable field theories on a ring \cite{Zam90} and on a strip with integrable boundaries \cite{LeC95}, including the sine-Gordon model with two boundaries \cite{Cau03,Ahn04}. In those problems the interaction is localized at the end points of the interval. Here it extends over the two half-lines outside of the interval, where it produces a bulk gap, so that the length scale $\xi$ enters via the finite penetration depth into the gapped regions rather than via a local boundary coupling. A sine-Gordon model with the cosine switched off in a window is not integrable, as far as we know, and it is only at $K=1/5$ that we have the full crossover function.

The mapping at $K=1/5$ onto a Josephson junction complements the analogy between anomalous reflection and Andreev reflection \cite{Ued26}. Here the mapping is not a relabeling of the fermions: the fermionic field $\Psi_j$ is a soliton, nonlocal in terms of the original chiral fermions, and the conserved charges $Q,Q'$ correspond to the spin of the superconducting problem, not to its charge.

We conclude with a direction for future work. The 3-4-5-0 model has a second, inequivalent, symmetric gapping interaction, and a junction between unlike gapped regions binds a Majorana zero mode \cite{Boy20}. In the long-junction regime one of the two bosonic fields then has a Dirichlet boundary condition at one interface and a Neumann condition at the other, which changes the phase-independent Casimir energy from $-\pi\hbar v/12L$ to $-\pi\hbar v/48L$. It would be of interest to follow that junction into the short-junction regime. The Luther-Emery mapping that we have used here does not directly extend to a junction between unlike gapped regions.

\acknowledgments

I am indebted to Atsushi Ueda and Vladimir Zakharov for explaining to me the anomalous reflection process at a symmetrically gapped interface. My research is supported by the Netherlands Organisation for Scientific Research (NWO/OCW), as part of Quantum Limits (grant ID 10.61686/EBSOG21615).  I used Claude Opus/Fable and GPT Astra as an interactive tool to explore the topic. Access to these large language models was provided through OpenAI's ChatGPT for Academic Researchers program and Anthropic's Claude Team plan for scientists. 

\appendix

\section{Perturbation theory for the short junction}
\label{app:pert}

We consider one of the two copies ${\cal H}_j$ of the bosonized Hamiltonian \eqref{eq:HLL2} and drop the index $j$ for ease of notation. The velocity is $v$, the Luttinger parameter is $K$, and the scaling dimension is $d=5K$.

We split the Hamiltonian \eqref{eq:HLL} as $H=H_0+V$, with $H_0$ the quadratic part (a gapless Luttinger liquid on the whole line, with uniform $K$) and
\begin{equation}
\begin{split}
&V=V_-+V_+,\\
&V_\pm=-g_0\int_{\pm x>L/2}dx\,\cos\bigl(2\sqrt{5}\,\phi(x)-\alpha^\pm\bigr),
\end{split}
\end{equation}
the cosine potential in the left and right region.

In the absence of the cosine potential, the field $\phi(x,\tau)$ is a 1+1 dimensional Gaussian field in space $x$ and imaginary time $\tau$. The free-field correlator is \cite{Giamarchi}
\begin{equation}
\begin{split}
&\langle[\phi(x,\tau)-\phi(x',0)]^2\rangle_0=K\ln(r/a_0),\\
&r=\sqrt{(x-x')^2+(v\tau+a_0)^2},
\end{split}
\end{equation}
and Gaussian averaging of the exponentials gives
\begin{align}
\langle e^{2i\sqrt{5}\phi(x,\tau)}e^{-2i\sqrt{5}\phi(x',0)}\rangle_0={}&e^{-10\langle[\phi(x,\tau)-\phi(x',0)]^2\rangle_0}\nonumber\\
={}&(a_0/r)^{2d}.
\end{align}

Since the free Hamiltonian is invariant under a constant shift of the $\phi$-field, $\langle e^{i\beta\phi}\rangle_0=0$ and also $\langle e^{i\beta\phi(x,\tau)}e^{i\beta\phi(x',0)}\rangle_0=0$. Therefore, upon averaging of a product of cosines, only the exponents with opposite sign contribute,
\begin{align}
&\langle\cos(2\sqrt{5}\phi(x,\tau)-\alpha)\cos(2\sqrt{5}\phi(x',0)-\alpha')\rangle_0=\nonumber\\
&\qquad\qquad\tfrac{1}{2}\cos(\alpha'-\alpha)(a_0/r)^{2d}.
\end{align}

This is the input needed for a perturbative calculation of the energy correction $\delta E$ to the free ground state energy,
\begin{align}
\delta E=-\frac{1}{\hbar}\int_0^\infty d\tau\,\langle V(\tau)V(0)\rangle_0 +{\cal O}(V^3),
\end{align}
where we have used that $\langle V\rangle_0=0$. Substitution of $V=V_++V_-$ gives four terms, of which only the cross terms $V_+V_-$ and $V_-V_+$ depend on $L$. Their contribution to the energy is
\begin{align}
{\cal E}_{\rm C}={}&-\frac{1}{\hbar}g_0^2\cos(\alpha'-\alpha)\int_{-\infty}^{-L/2}dx\int_{L/2}^{\infty}dx'\int_{0}^\infty d\tau\nonumber\\
&\times\Bigl(\frac{a_0^2}{(x'-x)^2+(v\tau+a_0)^2}\Bigr)^{d}.\label{ECcross}
\end{align}
The integral over $x$ and $x'$ converges at large distances only for $d>3/2$. For smaller $d$ it is cut off at distances of order $\xi_{\rm eff}$, where the free correlator no longer applies because the cosine potential pins the field. The derivative with respect to $L$ does not depend on that cutoff for $d>1$, so we take the derivative before removing the cutoff. This gives the force (with $\alpha'-\alpha=\delta$)
\begin{align}
{\cal F}_{\rm C}={}&-\frac{1}{\hbar}g_0^2\cos(\alpha'-\alpha)\int_{0}^{\infty}ds\int_{0}^\infty d\tau\nonumber\\
&\times\Bigl(\frac{a_0^2}{(L+s)^2+(v\tau+a_0)^2}\Bigr)^{d}.
\end{align}

The integral over imaginary time is convergent for $d>1/2$, when it evaluates to
\begin{align}
{\cal F}_{\rm C}={}&-\frac{a_0^{2d}}{\hbar v}g_0^2\cos(\alpha'-\alpha)\int_0^\infty ds\,(L+s)^{1-2d}\nonumber\\
&\times\frac{\sqrt{\pi}\Gamma(d-1/2)}{2\Gamma(d)}[1+{\cal O}(a_0/L)].
\end{align}
The integral over $s$ requires $d>1$ for convergence, resulting in
\begin{equation}
{\cal F}_{\rm C}=\frac{a_0^{2d}}{\hbar v}g_0^2\cos(\alpha'-\alpha)\frac{\sqrt{\pi}\Gamma(d-1/2)}{{2\Gamma(d)}}\frac{L^{2-2d}}{2-2d}.\label{Fcasymptotic}
\end{equation}
This is Eq.\ \eqref{eq:Fpert}, upon summing over the two independent Luttinger liquids.

The expansion parameter is the coupling at the scale $L$, $g_0a_0^dL^{2-d}/\hbar v\simeq(L/\xi_{\rm eff})^{2-d}$, which is small in a short junction because the cosine is a relevant perturbation. The free correlator applies up to distances of order $\xi_{\rm eff}$. Cutting off the integral over $s$ at that distance changes the force by an $L$-independent term of order $\xi_{\rm eff}^{2-2d}$, which is subleading for $d>1$.

\section{State space of the refermionized model}
\label{app:sectors}

The mapping of Sec.\ III relates the refermionized 3-4-5-0 model at $K_1=K_2=1/5$ to two Dirac fermions $\Psi_1,\Psi_2$ in the BCS Hamiltonian. The latter fields are not local in terms of the original chiral fermions $\Psi_q$, $q\in\{3,4,5,0\}$. The state space may therefore be different despite the identical Hamiltonian densities \eqref{operatoridentity}. This is the distinction made by Klassen and Melzer \cite{Kla93} when comparing the mapping of the sine-Gordon model onto the massive Thirring model \cite{Col75}. Here we identify the common sectors of the 3-4-5-0 and BCS Hamiltonians, and show that they suffice to determine the Casimir energy of the 3-4-5-0 junction from the known results for the SNS junction \cite{Bee23}.

\subsection{Soliton number and chiral number}

To have a discrete set of states we close the system into a ring of circumference $R$, which we will send to infinity at the end. We choose antiperiodic boundary conditions.  The chiral bosonic fields $\varphi_q$ count the fermions of species $q$. The occupation number (relative to the filled Fermi sea) is the winding number
\begin{equation}
n_q=\frac{s_q}{2\pi}\oint dx\,\partial_x\varphi_q\in\mathbb{Z},\label{nqwinding}
\end{equation}
with $s_q=+1$ for the right-movers ($q=3,4$) and $s_q=-1$ for the left-movers ($q=5,0$). The four integers $n_q$ are independent.

The transformed fields $\eta_{{\rm R},j},\eta_{{\rm L},j}$ define integrated right-moving and left-moving densities of $\Psi_j$,
\begin{equation}
N_{{\rm R}j}=\frac{1}{2\pi}\oint dx\,\partial_x\eta_{{\rm R}j},\;\;N_{{\rm L}j}=-\frac{1}{2\pi}\oint dx\,\partial_x\eta_{{\rm L}j}.
\end{equation}
At $K_j=1/5$ one has $\eta_{{\rm R}j}-\eta_{{\rm L}j}=2\sqrt5\,\phi_j$ and $\eta_{{\rm R}j}+\eta_{{\rm L}j}=2\theta_j/\sqrt5$, so the soliton number $N_j=N_{{\rm R}j}+N_{{\rm L}j}$ and the chiral number $J_j=N_{{\rm R}j}-N_{{\rm L}j}$ are given by
\begin{equation}
N_j=\frac{\sqrt5}{\pi}\oint dx\,\partial_x\phi_j,\;\;J_j=\frac{1}{\sqrt5\,\pi}\oint dx\,\partial_x\theta_j.\label{NJdef}
\end{equation}
Substitution of Eqs.\ \eqref{phivarphi} and \eqref{nqwinding} gives
\begin{equation}
\begin{split}
&N_1=n_3-2n_4-n_5-2n_0,\\
&N_2=2n_3+n_4+2n_5-n_0,\\
&J_1=\tfrac{1}{5}(n_3-2n_4+n_5+2n_0),\\
&J_2=\tfrac{1}{5}(2n_3+n_4-2n_5+n_0).
\end{split}\label{NJn}
\end{equation}

The soliton numbers are integers. The combinations $-N_1+2N_2=3n_3+4n_4+5n_5$ and $2N_1+N_2=4n_3-3n_4-5n_0$ are the charges $Q$ and $Q'$ of Eq.\ \eqref{eq:charge}, and $N_1+N_2=\sum_q n_q\bmod 2$, so the fermion parities of the two models match. The elementary soliton charge of field $j=1$ can be carried by a composite with occupation changes $(n_3,n_4,n_5,n_0)=(1,-1,0,1)$, with charge $(Q,Q')=(-1,2)$; that of field $j=2$ by a composite with occupation changes $(1,1,-1,0)$, with charge $(2,1)$. All pairs $(N_1,N_2)\in\mathbb{Z}^2$ can therefore be realized.

\subsection{Comparison of the two models}

The chiral numbers $J_j$ are multiples of $1/5$. Elimination of the $n_q$'s from Eq.\ \eqref{NJn} shows that the quantum numbers that are realized in the 3-4-5-0 model are
\begin{equation}
\begin{split}
&J_1\in N_1-\tfrac{2}{5}N_2+2\mathbb{Z},\\
&J_2\in N_2+\tfrac{2}{5}N_1+2\mathbb{Z}.
\end{split}\label{lattice3450}
\end{equation}
For the Dirac fermions, on the other hand, $N_{{\rm R}j}$ and $N_{{\rm L}j}$ are integers, hence
\begin{equation}
J_j\in N_j+2\mathbb{Z}.\label{latticeDirac}
\end{equation}
At fixed $N_1,N_2$, the two lattices of chiral occupations coincide if and only if $N_1$ and $N_2$ are both multiples of 5: for those soliton numbers Eq.\ \eqref{lattice3450} reduces to $J_j\in N_j+2\mathbb{Z}$.

An offset of $J_j$ can be represented by a twisted boundary condition,
\begin{equation}
\begin{split}
&\psi_{{\rm R/L},j}(x+R)=-e^{i\vartheta_j}\psi_{{\rm R/L},j}(x)\\
&\Rightarrow J_j=J_j^{(0)}+\vartheta_j/\pi.
\end{split}
\end{equation}
Eq.\ \eqref{lattice3450} is therefore reproduced by
\begin{equation}
\vartheta_1=-\frac{2\pi}{5}N_2,\;\;
\vartheta_2=+\frac{2\pi}{5}N_1.
\label{mutualtwists}
\end{equation}
The mass bilinear $\psi_{{\rm R}j}^{\dagger}\psi_{{\rm L}j}^{\vphantom{\dagger}}$ is single-valued under this twist, since the two chiral components acquire the same phase. The independent Dirac model has $\vartheta_1=\vartheta_2=0$.

For a single field Ref.\ \onlinecite{Kla93} distinguishes the lattice $J\in2\mathbb Z$ of the sine-Gordon model from the lattice $J\in N+2\mathbb Z$ of the massive Thirring model. Here the two-field lattice instead produces a $\mathbb Z_5$ twist of one field determined by the soliton number of the other. The same $\mathbb Z_5$ twist features prominently in studies of the anomalous reflection of a chiral fermion from an SMG-gapped region \cite{vBe25}.

\subsection{Local composites and soliton strings}

The nonlocality of the refermionized fields can be displayed explicitly. Substitution of Eq.\ \eqref{phivarphi} into Eq.\ \eqref{refermionization} gives the vertex-operator identities
\begin{equation}
\begin{split}
\Psi_3^{\vphantom{\dagger}}\Psi_4^\dagger\Psi_0^{\vphantom{\dagger}}
 &\propto\psi_{{\rm R}1}W_2,\;\;
\Psi_4^{\vphantom{\dagger}}\Psi_5^\dagger\Psi_0^\dagger
 \propto\psi_{{\rm L}1}W_2,\\
\Psi_3^{\vphantom{\dagger}}\Psi_4^{\vphantom{\dagger}}\Psi_5^\dagger
 &\propto\psi_{{\rm R}2}W_1^\dagger,\;\;
\Psi_3^\dagger\Psi_5^{\vphantom{\dagger}}\Psi_0^\dagger
 \propto\psi_{{\rm L}2}W_1^\dagger.
\end{split}\label{composites}
\end{equation}
The string fields $W_j$ are defined by
\begin{equation}
\begin{split}
&W_j(x)=e^{2i\phi_j(x)/\sqrt5},\\
&W_j(x)=W_j(x_0)\exp\left[\frac{2\pi i}{5}
\int_{x_0}^{x} n_j(x')\,dx'\right],
\end{split}\label{Wstring}
\end{equation}
where $n_j=(\sqrt5/\pi)\partial_x\phi_j$ is the soliton density and $x_0$ is a reference point. Thus a local three-fermion composite differs from a Dirac field by a string counting the solitons of the other field, with a phase $2\pi/5$ per soliton. The strings cancel from the mass bilinears. On the ring,
\begin{equation}
W_j(x+R)=e^{2\pi iN_j/5}W_j(x),
\end{equation}
which accounts for the relative twists in Eq.\ \eqref{mutualtwists} when the composites obey the original antiperiodic boundary conditions.

\subsection{Casimir energy}

The Hamiltonian \eqref{eq:HLL} contains the densities $\partial_x\phi_j$, $\partial_x\theta_j$, and the gapping vertices $\exp(\pm2i\sqrt5\,\phi_j)$. The densities preserve $N_j$ and $J_j$. The vertices preserve $N_j$ and shift $J_j$ by $\pm2$, just as the Dirac mass bilinear converts a left-mover into a right-mover or vice versa. In the original variables the two interactions change the occupations by
\begin{equation}
\begin{split}
\boldsymbol\ell_1&=(1,-2,1,2),\\
\boldsymbol\ell_2&=(2,1,-2,1),
\end{split}
\end{equation}
or their negatives. Eq.\ \eqref{NJn} shows that each changes only the corresponding $J_j$ by $\pm2$. At fixed $N_1,N_2$, the offsets of $J_1,J_2$ modulo two are therefore preserved.

In a common sector, the allowed zero modes and the oscillator Fock spaces coincide. The gapping vertices have identical matrix elements in these spaces, including their shifts of $J_j$ by two. The Hamiltonians consequently have identical matrix representations in that sector.

In particular, the sector $N_1=N_2=0$ has $J_1,J_2\in2\mathbb Z$ in both models and contains the unperturbed vacuum. Its original occupation vectors are precisely
\begin{equation}
(n_3,n_4,n_5,n_0)=m_1\boldsymbol\ell_1+m_2\boldsymbol\ell_2,
\;\; m_1,m_2\in\mathbb Z,
\end{equation}
with $J_j=2m_j$. The two models therefore have the same spectrum in this neutral sector for any mass profile, including any junction length $L$ and phase differences $\delta_j$, on the finite ring with the boundary conventions specified above.

To recover the ground state Casimir energy, we use the finite-$D$ geometry of the main text. The ring contains the junction of length $L$, two gapped segments of length $D$, and an outer gapless region of length
\begin{equation}
S=R-L-2D.
\end{equation}
We first take $R\to\infty$ at fixed $L,D$, and then take $D/\xi\to\infty$ at fixed $L$. At finite $D$, the outer gapless region can accommodate a compensating charge at an energy cost of order $\hbar v_{\rm F}/S$. Thus fixing the total charges to zero does not constrain the equilibrium charge of the junction in the thermodynamic limit. In particular, the fermion parity of the junction may be even or odd, while keeping the global fermion parity of the entire ring even. We have thus established the Casimir energy correspondence of Eq.\ \eqref{E3450ESNS}.

\end{document}